\documentclass[runningheads]{llncs}
\usepackage{amsmath}
\usepackage{graphicx}
\usepackage{hyperref}
\usepackage{courier}
\usepackage{multicol}
\usepackage{hyperref}
\usepackage{academicons}
\usepackage{xcolor}
\usepackage{todonotes}
\usepackage{subfigure}
\usepackage{array}
\usepackage{multirow}
\usepackage{ulem}
\newcommand{\orcid}[1]{\href{https://orcid.org/#1}{\textcolor[HTML]{A6CE39}{\aiOrcid}}}

\begin{document}
\title{AutoPET Challenge 2023: Sliding Window-based Optimization of U-Net}
%
\titlerunning{AutoPET: Combining nn-Unet, Swin UNETR, and MIP Classifiers}
%
\author{Matthias Hadlich\inst{1}$^{,*}$ \and
Zdravko Marinov\inst{1,2}$^{,*}$ \and
Rainer Stiefelhagen\inst{1}
}
\authorrunning{M. Hadlich et al.}
%
\institute{Institute for Anthropomatics and Robotics, Karlsruhe Institute of Technology, Karlsruhe, Germany \\ 
\email{matthias.hadlich@student.kit.edu}, \email{\{zdravko.marinov, rainer.stiefelhagen\}@kit.edu} 
\and HIDSS4Health - Helmholtz Information and Data Science School for Health, Karlsruhe/Heidelberg, Germany
}

\def\thefootnote{*}\footnotetext{Shared first author}\def\thefootnote{\arabic{footnote}}

\maketitle              
\begin{abstract}
Tumor segmentation in medical imaging is crucial and relies on precise delineation. Fluorodeoxyglucose Positron-Emission Tomography (FDG-PET) is widely used in clinical practice to detect metabolically active tumors. However, FDG-PET scans may misinterpret irregular glucose consumption in healthy or benign tissues as cancer. Combining PET with Computed Tomography (CT) can enhance tumor segmentation by integrating metabolic and anatomic information. FDG-PET/CT scans are pivotal for cancer staging and reassessment, utilizing radiolabeled fluorodeoxyglucose to highlight metabolically active regions. Accurately distinguishing tumor-specific uptake from physiological uptake in normal tissues is a challenging aspect of precise tumor segmentation. The AutoPET challenge addresses this by providing a dataset of 1014 FDG-PET/CT studies, encouraging advancements in accurate tumor segmentation and analysis within the FDG-PET/CT domain. \\
Code: \url{https://github.com/matt3o/AutoPET2-Submission/}
\end{abstract}
\section{Introduction}
\label{sec:introduction}
In the domain of oncological diagnostics, the integration of Fluorodeoxyglucose Positron-Emission Tomography (FDG-PET) and Computed Tomography (CT) has assumed a pivotal role, facilitating comprehensive insights into the metabolic dynamics of various malignant solid tumor entities \cite{ben200918f}. FDG-PET, acknowledged for its capacity to delineate glucose consumption within tissues, holds significant promise in therapy control and monitoring, owing to the characteristic escalated glucose uptake by tumor lesions \cite{egger2022medical}. However, the non-specificity of FDG-PET often introduces interpretational ambiguities, as it may also manifest in benign or healthy tissue \cite{gatidis2023autopet}, potentially leading to erroneous diagnoses.

To mitigate this diagnostic challenge, the fusion of PET with CT has emerged as an integrated approach, combining metabolic data with precise anatomical information. This combination enhances tumor detection accuracy \cite{ben200918f}, \cite{marinov2023mirror}, offering a cohesive synergy particularly valuable in clinical practice \cite{gatidis2023autopet}.

Within this evolving landscape of medical diagnostics, the Automatic Lesion Segmentation in Whole-Body FDG-PET/CT Challenge (AutoPET)\footnote{\url{https://autopet-ii.grand-challenge.org/}} embodies a critical juncture. It motivates researchers and practitioners to develop automated, bi-modal methodologies for the three-dimensional segmentation of tumor lesions embedded within  FDG-PET and CT scans \cite{gatidis2023autopet}. The challenge accelerates advancements in deep learning-based automated tumor lesion segmentation through the provision of a large densely annotated dataset of 1014 volumes.

In this work, we propose using the well-known U-Net architecture \cite{ronneberger2015u} to tackle the AutoPET challenge. Despite the ubiquity of U-Net models in medical segmentation tasks \cite{isensee2021nnu}, \cite{eisenmann2023winner}, achieving high performance in the domain of whole-body PET/CT lesion segmentation has remained elusive \cite{liu2022autopet}, \cite{zhong2022autopet}, \cite{heiliger2022autopet}, \cite{ye2022exploring}, \cite{hallitschke2023multimodal} largely due to the scarcity of training data in preceding studies \cite{egger2022medical}. Drawing upon the insights provided by the AutoPET Challenge U-Net-based winner from 2022 \cite{ye2022exploring}, we undertake a practical investigation to understand the important training parameters of the U-Net model for segmenting lesions. We believe that it is possible to achieve a better and more robust model by focusing on the intricacies of data pre-processing, data augmentation, learning rate scheduling, and crop-size selection during model training. 
Our work and model are based on prior experiments in interactive segmentation \cite{marinov2023guiding}. Thus, for our hyperparameter tuning experiments, we present results using our interactive model. Nonetheless, for our final submission, we exclude the integration of interactive clicks into the model and employ its optimal hyperparameter configuration.

\section{Methodology}
\subsection{Model Architecture}\label{sec:pipeline}

The model used for the challenge is called DynUNet, which is an adaption of the UNet for the MONAI library \cite{cardoso2022monai}.
Contrary to the default UNet,  DynUNet does not use max-pooling for downsampling but instead uses strided convolutions. 
Additionally, the residual is passed through a convolutional layer such that the input size from the downsampling layer matches the output size of this layer.
All of the changes can be traced back to three prior works: \cite{isensee2019automated}, \cite{isensee2018nnu}, and \cite{futrega2021optimized}.

\begin{figure}[h]
    \centering
    \caption{An overview of the used DynUNet architecture.}
    \includegraphics[width=1.0\textwidth]{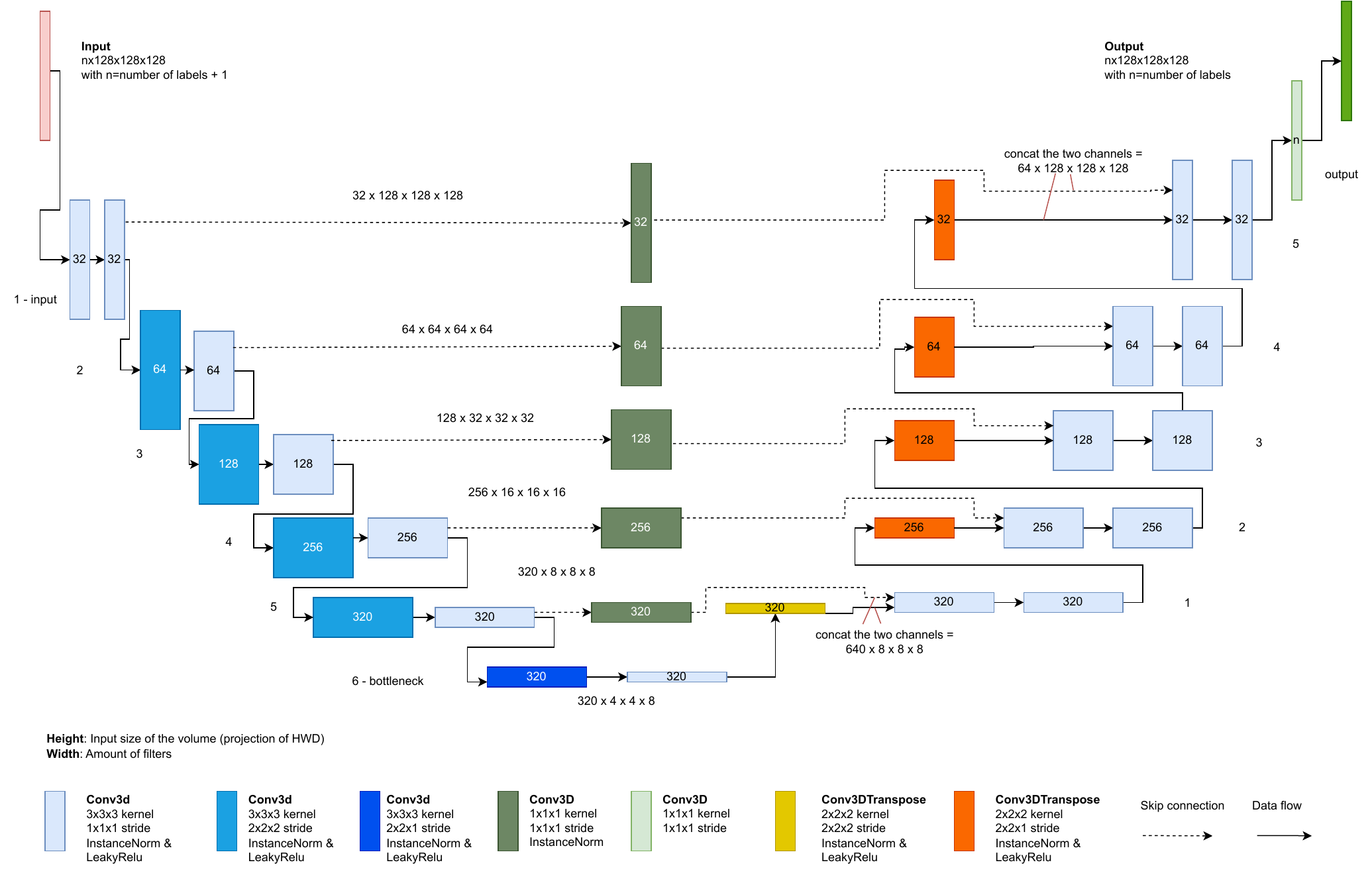}
    \label{fig:dynunet}
\end{figure}

Our default configuration of the network consists of six layers of filter size [\texttt{32, 64, 128, 256, 320, 320}].
As discussed above, the convolutions are strided with a size of [\texttt{1, 2, 2, 2, 2, [2, 2, 1]}], and the upsampling is thus done in the inverse order.
An architectural diagram can be found in Figure \ref{fig:dynunet}.

\subsection{Data Pre-processing and Augmentation}
\textbf{Pre-processing.} We restrict ourselves by using only the PET volumes from the paired PET/CT scans. We apply multiple pre-processing transformations to each batch of data. Apart from changing the channel order, the orientation is set to a RAS (Right-Anterior-Superiror) coordinate system. As the AutoPET spacing is $\approx$[2, 2, 3]mm$^3$, the data is resampled accordingly with this fixed voxel size.
The intensity of each PET image is scaled, based on its voxel intensity statistics, with MONAI's \texttt{ScaleIntensityRangePercentiled} to the 0.05 and 99.95 percentiles. During training, a random crop of size \texttt{224x224x224} is sampled, with a probability of 0.6 of being centered around a tumor lesion and 0.4 of being centered around the background. To achieve this, we utilize the \texttt{RandCropByPosNegLabeld} MONAI transform. This crop is balanced by the class label of the voxel in the crop's center -  in 60\% of the cases the voxel is positive, and in the other 40\% it is negative. This ensures that the network learns about positive and negative samples in a more balanced training regime.

\textbf{Data Augmentation.} We apply two types of data augmentation - random flipping and random rotation. We apply a random flip on each spatial axis with a probability of 0.1. We also apply a random 90-degree rotation with a probability of 0.1 for each axis. 

\subsection{Data Post-processing}

Since we are using a sliding window approach, the final prediction volume gets stitched from the various output patches.
This process is done with a user defined overlap, in our case this was set to 75\%.

After the result prediction a softmax is applied.

For the ensemble based solution the two steps mentioned above are done for each of the five networks prediction separately.
After the softmax on each prediction a voting mechanism combines the different predictions into a single one.

\subsection{Hyperparameter Tuning}

As explained above most of the different experiments have been run on interactive code.
Nevertheless they should be representative in terms of general performance of the network.
Variations of +/- 0.5\% Dice are to be expected since the guidance signal was non-deterministic.

\subsubsection{Sliding window versus normal inferer}

First of all we compare the sliding window infererence to the normal one, figure \ref{tab:sw_vs_simple}.
As it can be seen in the table, on the interactive code the sliding window inferer wins with a lead of 2.81\% Dice.

Next different region of interest sizes have been tried out.
The best performing one here was the \texttt{128x128x128} crop.
Note that the sliding window was active during training. 
In the thesis it is shown that training with overlap active gains about 1\% of Dice.

This overlap means for the \texttt{128x128x128} instead a window of size \texttt{320x320x320} has been calculated, with calculations being equal to a normal inferer of size \texttt{384x384x384}.
As we can see a lot of overhead calculations are being done by the sliding window inferer.
However in the next subsection we will show that the overhead calculations for the overlap actually lead to a better Dice score.

\begin{table}[h]
\label{tab:sw_vs_simple}
\centering
\caption{Interactive run of Sliding Window versus Simple Inferer}
    \begin{tabular}{|c|c|c|}
        \hline
        &\textbf{Sliding Window} & \textbf{Simple Inferer}\\
        \hline
        Dice & \textbf{83.83\%} & 81.02\%\\
        \hline
    \end{tabular}
\end{table}

\begin{table}[h]
\label{tab:intensity_scaling}
\centering
\caption{Different region of interest sizes compared. Trained on a crop of size \texttt{256x256x256}.}
    \begin{tabular}{|c|c|c|c|c|}
        \hline
        & \textbf{64x64x64} & \textbf{128x128x128} & \textbf{192x192x192} & \textbf{256x256x256}\\
        \hline
        Dice (validation) & 84.74\% & \textbf{85.22\%} & 83.66\% & 84.75\%\\
        \hline
        Dice (training) & 87.99\% & 88.46\% & \textbf{88.98\%} & 88.79\%\\
        \hline
    \end{tabular}
\end{table}

\subsubsection{Sliding window overlap}

Now we will look at the overlap of the sliding window inferer.
Table \ref{tab:sw_overlap} shows that increasing the overlap also increases the Dice score of the network.
In our experiments the higher the overlap the better the results have been.
This can be seen as a way of creating a mini Ensemble with same weights.
The overlap uses a Gaussian fade away to make the regions closer the center weight more heavily when stitching together the final output.

Additionally experiments have been run to verify the impact of training with overlap on.
Table \ref{tab:sw_overlap_0}, which shows a network trained on 0\% overlap, overall shows slightly worse results, especially for the higher overlaps it becomes significant.
As expected running it with 0 overlap returns slightly better results than the network trained with overlap being forced to use none.
We can thus conclude that activating overlap during training enhances the final score.

\begin{table}[h]
\label{tab:sw_overlap}
\centering
\caption{Non-interactive validation runs with different settings for the overlap. The network has been trained on 25\% overlap.}
\begin{tabular}{|c|c|c|}
    \hline
    \textbf{Experiment} & \textbf{Overlap} & \textbf{Dice} \\
    \hline
    201 & 0 & 66.33\% \\
    202 & 0.25 & 73.04\% \\
    203 & 0.5 & 73.54\% \\
    207 & 0.75 & \textbf{74.07\%} \\
    \hline
\end{tabular}
\end{table}

\begin{table}[h]
\centering
\caption{Non-interactive validation runs with different settings for the overlap. The network has been trained on 0\% overlap.}
\label{tab:sw_overlap_0}
\begin{tabular}{|c|c|c|}
    \hline
    \textbf{Experiment} & \textbf{Overlap} & \textbf{Dice} \\
    \hline
    v\_208\_0.0 & 0 & 66.57\% \\
    v\_208\_0.25 & 0.25 & 71.35\% \\
    v\_208\_0.5 & 0.5 & 71.99\% \\
    v\_208\_0.75 & 0.75 & \textbf{72.86}\% \\
    \hline
\end{tabular}
\end{table}

\subsubsection{Convergence behaviour with different losses}

\begin{figure}[hb!]
    \centering
    \caption{Comparing the Dice Loss, in MONAI called \texttt{MeanDice} to the \texttt{DiceCELoss}.}
    \subfigure[Validation Dice of run \texttt{158}. In this run the \texttt{DiceCELoss} has been used as the loss.]{\includegraphics[width=0.45\linewidth]{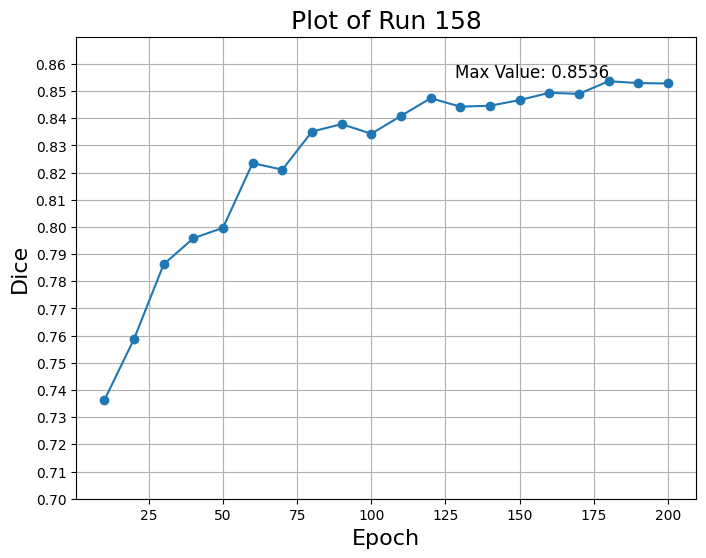}}
    \hspace{0.05\textwidth}
    \subfigure[Validation Dice of run \texttt{183}. In this run the \texttt{MeanDice} has been used as the loss.]{\includegraphics[width=0.45\linewidth]{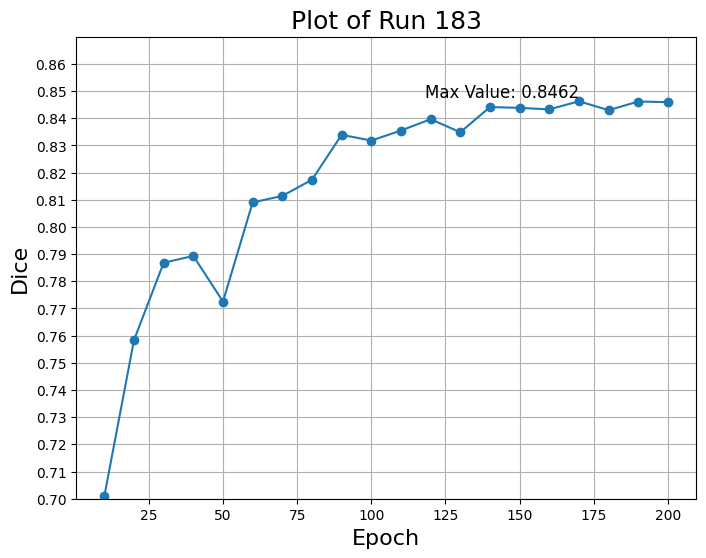}}
\label{fig:dicevsdiceceloss}
\end{figure}

Figure \ref{fig:dicevsdiceceloss} shows the convergence behaviour of the Dice loss vs the DiceCELoss.
As it can be seen the DiceCELoss start with a higher initial validation Dice in epoch 10, 73.62\% against 70.09\%.
Also the final Dice metric was a little bit higher, 85.47\% for DiceCELoss and 84.62\% for Dice loss.
However a plateau appears to be reached for both losses.
In other experiments with more iterations it was shown that this method can reach a validation Dice of up to 87.60\%.

We can thus fully recommend the DiceCELoss as a standard choice for training.
It converges faster and also yields higher final scores especially in terms of Dice.

\subsubsection{Intensity scaling options}

Finally a quick comparison of different intensity scaling options.
The base run was a pre-calculated batch statistics normalization to the 0.005 and 99.95 percentiles of the intensity.
The first ScaleIntensityRangePercentiled applied the same percentiles but this time based on the statistics of a each item.
The last ScaleIntensityRangePercentiled is a base run with no clipping of the intensities, it only normalizes the intensity from 0 to 1.

As we can see the item-wise statistics outperformed the batch-wise statistics and the clipless method.

\begin{table}[h]
\label{tab:intensity_scaling}
\centering
\caption{Different ScaleIntensity settings compared.}
    \begin{tabular}{|c|c|c|c|}
        \hline
        & \textbf{Base run} & \textbf{ScaleIntensity-} & \textbf{ScaleIntensity-}\\
        & \textbf{CosineAnnealingLr} & \textbf{RangePercentiled} & \textbf{RangePercentiled 2}\\
        & (104) & (148) & (149)\\
        \hline
        Dice & 85.63\% & \textbf{86.69\%} & 85.44\%\\
        \hline
    \end{tabular}
\end{table}

\begin{table}[h]
    \centering
    \caption{Number of training epochs for each cross-validation model in our final submission of an ensemble and its corresponding best Train Dice.}
    \begin{tabular}{|c||c|c|c|}
    
    \hline
        Model & \# Epochs & Best Train Dice & Notes \\ \hline
        \sout{1} &  \sout{414} & \sout{83.92\%} & Network did not get included due to NaN errors \\
        2 & 564 & 87.33\% & \\
        3 & 447 & 85.92\% & \\
        4 & 411 & 85.44\% & \\
        5 & 391 & 86.76\% & \\ \hline
    \end{tabular}

    \label{tab:checkpoints}
\end{table}

\subsubsection{Best parameters}

A summary of the best found results can be found in table \ref{tab:best_settings}.

\begin{table}[h]
\label{tab:best_settings}
\centering
\caption{Best settings}
    \begin{tabular}{|c|>{\centering\arraybackslash}p{9cm}|}
        \hline
        \textbf{Parameter name} & \textbf{Setting}\\
        \hline
        Network & \texttt{DynUNet} with \texttt{[32, 64, 128, 256, 320, 320]} filters and a  depth of sex layers\\
        \hline
        Loss & \texttt{DiceCELoss} with \texttt{squared\_pred=True} \\
        & \texttt{and include\_background=True} \\
        \hline
        Optimizer & Adam \\
        \hline
        Learning rate scheduler & \texttt{CosineAnnealingLR} (\texttt{initial lr}=\texttt{2e-4}, \texttt{eta\_min=1e-8}) \\
        \hline
        Inferer & Sliding window inferer with ROI size \texttt{128x128x128}, sliding window overlap 0.75\\
        \hline
        Intensity scaling with & Custom Scaling to 0.05\% and 99.95\% intensity percentiles using \texttt{ScaleIntensityRanged}  \\
        \hline
        Automatic Mixed Precision & Active\\
        \hline
    \end{tabular}
\end{table}

\section{Proposed solutions to the AutoPET2 Challenge}

We propose two different approaches for the challenge as final submissions:

\begin{itemize}
    \item A single network with six layers as stated above, trained for 400 epochs.
    \item An ensemble of five networks, each with same six layers.
    Four of the networks were trained with cross-validation on five splits of the data. 
    The network trained on the first split did not get included, since it ran into NaN errors very quickly.
    They were trained without using the validation split for 800 epochs with no validation runs.
    However, all of the five networks did not finish in time for the challenge, so the most recent checkpoint was picked instead, where Table \ref{tab:checkpoints} summarizes how many epochs each model was trained for.
    
    Additionally, the best-performing single network was integrated as a teacher, bringing the total to five networks working collaboratively.
    The results of the different networks got combined with an equally weighted voting mechanism.
    
\end{itemize}

\newpage

\section{Results}
The results of our two final submissions can be seen in Table \ref{tab:your_transposed_table_label}. 
The Dice score is similar in both approaches but the FPV is significantly reduced in the ensemble, perhaps due to the smoothening effect on the predictions which filters outliers outside of the object. 
However, the single network has a much lower FNV, signifying a higher sensitivity to tumor detection.

\begin{table}[h]
    \centering
    \begin{tabular}{|l|l|l|l|l|}
        \hline
        \textbf{Method} & \textbf{Train dice} & \multicolumn{3}{c|}{\textbf{Results on the preliminary test set}} \\
        \hline
        & & Dice score & False negative volume & False positive volume\\
        \hline
        Single network & 86.76\% & 56.52\% & 0.0249 & 1.8015 \\
        \hline
        Ensemble & 86.16\% & 56.60\%&  	0.0572 & 1.0475\\
        \hline
    \end{tabular}
    \caption{The results of our method in the AutoPET2 challenge.}
    \label{tab:your_transposed_table_label}
\end{table}

\section{Post mortem: NaN errors during training if AMP is active}

In the preparation for the challenge we ran into NaN errors when training on A100 GPUs, but only when automated mixed precision was on.
During the debugging we found out our input already contained NaNs.

The reason in our case was a training crop to positive / negative areas of size \texttt{224x224x224}.
At the borders of the volume this resulted in crops which contained almost only 0s or even only 0s.
Our current hypothesis is that the normalization on the crop produces division by 0 errors.
This would make especially sense for the intensity scaling which might degrade if the input tensor only contains 0s.
However more debugging is necessary to find out the exact transform which produces the NaN errors.

The solution is to add a filter after the pre-transform to remap all NaN values to 0.
In our case this fixed the problem and we could resume training with AMP on the A100 GPUs.

\section{Acknowledgment}
The present contribution is supported by the Helmholtz Association under the joint research school “HIDSS4Health – Helmholtz Information and Data Science School for Health“. This work was performed on the HoreKa supercomputer funded by the
Ministry of Science, Research and the Arts Baden-Württemberg and by
the Federal Ministry of Education and Research.

%
%

%
%
%
%
\bibliographystyle{splncs04}
\bibliography{bib/refs}

\end{document}